\documentclass{article}

\usepackage{PRIMEarxiv}

\usepackage[utf8]{inputenc} 
\usepackage{amsmath}
\usepackage{hyperref}
\usepackage[T1]{fontenc}    
\usepackage{hyperref}       
\usepackage{booktabs}       
\usepackage{amsfonts}       
\usepackage{nicefrac}       
\usepackage{microtype}      
\usepackage{lipsum}
\usepackage{fancyhdr}       
\usepackage{graphicx}       
\graphicspath{{media/}}     

\title{Deep Learning Assisted Robust Detection Techniques for a Chipless RFID Sensor Tag
}

\author{
  Nadeem Rather*, Roy B. V. B. Simorangkir, John L. Buckley, Brendan O’Flynn, Salvatore Tedesco \\
  Tyndall National Institute\\
  University College Cork, Lee Maltings Complex, Dyke Parade, T12R5CP\\
  Cork, Ireland\\
  \texttt{*nadeem.rather@tyndall.ie} \\
}

\begin{document}
\maketitle

\begin{abstract}
In this paper, we present a new approach for robust reading of identification and sensor data from chipless RFID sensor tags. For the first time, Machine Learning (ML) and Deep Learning (DL) regression modelling techniques are applied to a dataset of measured Radar Cross Section (RCS) data that has been derived from large-scale robotic measurements of custom-designed, 3-bit chipless RFID sensor tags. The robotic system is implemented using the first-of-its-kind automated data acquisition method using an ur16e industry-standard robot. A large data set of 9,600 Electromagnetic (EM) RCS signatures collected using the automated system is used to train and validate four ML models and four 1-dimensional Convolutional Neural Network (1D CNN) architectures. For the first time, we report an end-to-end design and implementation methodology for robust detection of identification (ID) and sensing data using ML/DL models. Also, we report, for the first time, the effect of varying tag surface shapes, tilt angles, and read ranges that were incorporated into the training of models for robust detection of ID and sensing values. The results show that all the models were able to generalise well on the given data. However, the 1D CNN models outperformed the conventional ML models in the detection of ID and sensing values. The best 1D CNN model architectures performed well with a low Root Mean Square Error (RSME) of 0.061 (0.87\%) for tag ID and 0.0241 (3.44\%) error for the capacitive sensing. 
\end{abstract}


\section{Introduction}
Conventional passive Radio Frequency Identification (RFID) systems have seen widespread adoption in various industries, revolutionizing supply chain management, contact-less payments, asset-tracking, authentication, animal identification and sensing for various Internet of Things (IoT) applications \cite{finkenzeller2010rfid,slitr,catar,catar2,pigeon2021nfc,peres2020theoretical,gawade2019battery}. However, recent advances have paved the way for chipless RFID (CRFID), which eliminates the need for an integrated circuit silicon device component and offers cost-effectiveness, flexibility, sensing capabilities, as well as enhanced security \cite{5,6,7,8,karm}. In CRFID systems, Frequency Domain (FD) and Time Domain (TD) approaches are the two main techniques used for encoding and decoding information without the use of a silicon integrated circuit component.
In the FD technique, the information is encoded by manipulating the resonant frequencies or spectral characteristics of a tag antenna. The reader device transmits an EM wave and analyzes the frequency response of the tag using the backscattered EM wave to extract the encoded information. On the other hand, in TD approaches, the focus is on the examination of the temporal characteristics of the tag. Information encoding is accomplished by modulating the time delay or phase shift of the tag’s response to the interrogating signal from the reader \cite{7,9}. The FD techniques, particularly using Radar Cross Section (RCS), have been extensively explored due to their low design complexity for the development of CRFID tags \cite{7,10}. RCS-based CRFID tags encode data by modifying the reflective pattern of the tag antenna through variations of the antenna's shape, size, or material properties. 
\par The encoded data can be retrieved from the scattered Electromagnetic (EM) fields that the reader receives. However, the reading methods in CRFID systems have limitations when it comes to scalability, adaptability, and robustness. For accurate detection of the encoded data, different encoding techniques may require specific hardware configurations, antenna designs, or signal processing algorithms \cite{11}. This specificity can make it challenging to deploy CRFID systems in various applications and environments without extensive customization and optimization efforts. 
To reduce the reading complexity, several studies have leveraged the pattern recognition capabilities of both Machine Learning (ML) and its subfield, Deep Learning (DL), to effectively detect different EM signatures and retrieve accurate tag information \cite{12,13,14,15,16}. \par
As compared to conventional signal processing methods \cite{17,18}, the ML/DL approach offers a more flexible and adaptable solution for CRFID systems. By leveraging ML/DL algorithms and models, the CRFID systems can be designed to learn and adapt to different encoding techniques without the need for explicit rule-based programming or hardware modifications. The ML/DL algorithms can analyze the complex patterns and relationships within the CRFID data and extract meaningful information. These algorithms can also handle variations in signal strength, noise, and interference, leading to improved detection and identification accuracy. \par
As shown in \cite{12}, the authors use an RCS-based CRFID tag for hand gesture recognition. The tag is developed using three split-ring resonators to create a specific 3-bit EM signature in the backscattered EM wave. A Feed Forward Artificial Neural Network (ANN) is utilised for training the models and for successfully classifying the eight possible combinations. 
In \cite{13}, the authors utilize a DL-based security model to provide a high accuracy of greater than 93\% when classifying a cloned tag from a genuine CRFID tag, even in the presence of additive RF interference in real-time. 
In \cite{14}, the authors use RCS-based tags to develop a segregation system for a plastic  recycling application. A Random Forest classifier is trained with a dataset consisting of 300 data points to identify two plastic types (2 IDs). The authors achieved an accuracy of 90\% when classifying the two plastic types from homogenous bales. In another scenario involving non-homogeneous bales, an accuracy of 65\% was attained. In \cite{15}, the authors propose a comprehensive workflow for identification applications using ML. Three datasets are utilised for the implementation of ML models. When classifying a 4-bit CRFID tag, two datasets consisting of 2,400 instances with 600 measurements per tag and 5,600 instances with 900 measurements per tag are utilized to achieve a 100\% accuracy rate. 
These studies have achieved impressive results in classifying CRFID tags based on their EM signatures and RCS properties.\par
However, there are some important gaps in the existing literature that need to be addressed. Specifically, the impact of varying surface shapes on CRFID tag detection has not been extensively investigated, despite its relevance to real-world applications (such as supply chain management, asset tracking, healthcare and medical environments, where varying surface shapes can be encountered). In consideration of real-world applications, a simultaneous extraction of tag ID and sensing information from CRFID tags using ML approaches has not been explored either. Additionally, the focus on classification tasks leaves a gap in the exploration
of regression-based approaches for precise and continuous
prediction of tag ID and sensing information. Classification tasks identify the presence or absence of specific attributes in CRFID tags. However, it lacks the necessary granularity for certain real-world applications. For example, in supply chain management, merely knowing if a CRFID-tagged perishable item is "Within Required Temperature" or "Out of Required Temperature" is inadequate. Precise predictions of attributes like temperature and humidity are thus crucial for ensuring quality in such scenarios. In an effort to address the aforementioned gaps, this paper investigates:
\vspace{-0.5mm}
\begin{itemize}
    \item The impacts of varying tag surface shapes, orientations, and read ranges for the development of robust detection algorithms suitable for real-world implementation, aligning with the primary objective of incorporating ML/DL techniques in CRFID systems.
    \item The feasibility of extracting both ID and sensing information from EM signatures.
    \item The use of regression-based approaches for accurate and continuous prediction of tag IDs and sensing values.
    \item The utilization of a large dataset acquired through an automated data acquisition system, distinguishing it from prior studies.
\end{itemize}

Through these investigations, this paper contributes to a more comprehensive understanding and practical implementation of ML/DL-assisted CRFID systems. The paper is structured as follows. In Section II, we provide a detailed description of the CRFID sensor tag design and the automated data acquisition methodology. Section III focuses on the study of ML and DL-based modelling techniques. Moving forward to Section IV, we present the results obtained from these models, highlighting the best-performing techniques that are validated through a real-world demonstration. In Section V, we analyze the implications of our findings, compare them with previous research, and examine potential future improvements and research directions. Finally, in Section VI, we conclude the paper and discuss conceivable directions for future work.
\vspace{-2mm}
\section{Research Methodology}
\vspace{-1mm}
A single-layer 3-bit capacity RCS-based CRFID sensor tag was developed based on the design methodology discussed in \cite{6}. There are eight possible ID combinations with three sensing states associated with each ID. This results in a total of twenty-four sensing tags that need to be characterized. Furthermore, a first-of-its-kind automated data acquisition method using a robot was developed and utilized to obtain a large dataset for ML/DL implementation.
\vspace{-1mm}
\subsection{3-Bit Chipless RFID Sensor Tag}
\begin{figure}[t]
  \begin{center}
  \includegraphics[width=4.5in]{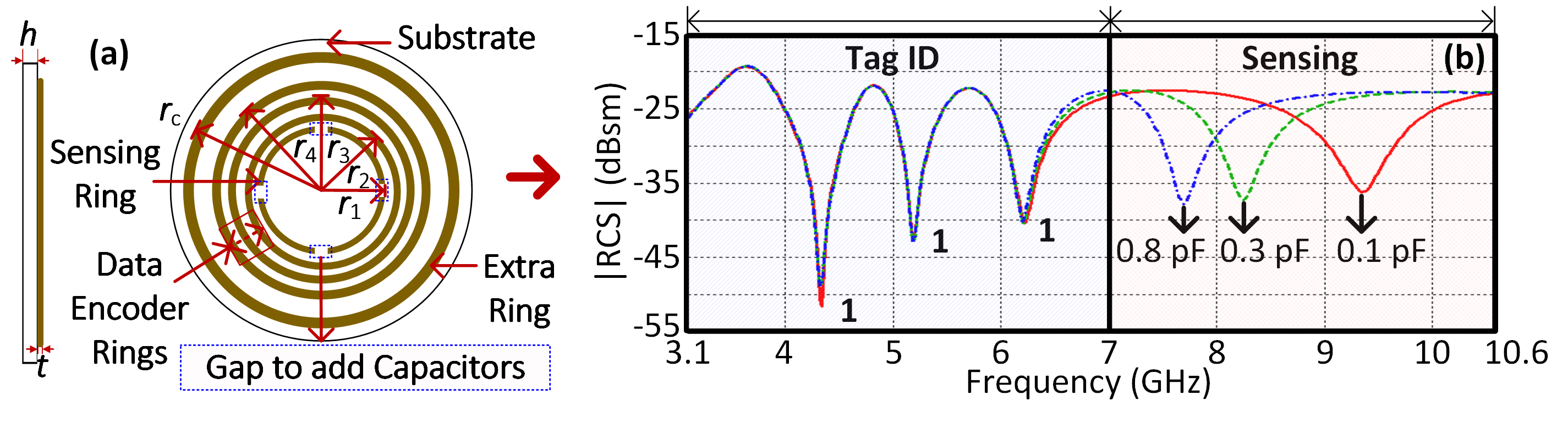}\\

 \caption{A 3-bit CRFID sensor tag: (a) Topology (Dimensions in millimeters are: $r_{1}$ = ~6, $r_{2}$ = ~7.12, $r_{3}$ = ~8.45, $r_{4}$ = ~10.03, $r_{c}$ = ~14.13, ${h}$ = ~0.125 and ${t}$ = ~0.01) (b) Simulated EM RCS response of tag ID 7 with 0.1, 0.3 and 0.8 pF sensing points.}
  \label{tag}
  \end{center}

\end{figure}

\vspace{-1mm}
A 3-bit polarization-insensitive capacitive sensing RCS-based CRFID tag is developed using nested circular ring resonators. The tag is fabricated through a screen printing process with a polyester mesh screen (300M) of an elastic silver ink WIK21285-89A from Henkel (conductivity = 3.94 x 10$^6$ S/m) on a flexible polyethylene terephthalate (PET) substrate ($\varepsilon_r$ = 2.9 and tan$\delta$ = 0.0025). The printed tag was sintered in an oven at $120^\circ$ for 15 minutes. The tag consists of an outermost ring to enable the null encoding, an innermost ring for sensing, and three rings in between for 3-bit ID data encoding (Fig.\ref{tag}). The procedure of tag design is described in \cite{6}, and optimized dimensions are provided in the caption of Fig.\ref{tag}. For the sensing feature, materials with stimuli-sensitive permittivity are to be added to the innermost ring, altering the ring's resonance position upon exposure to appropriate stimuli (e.g., humidity, temperature, and light, to name a few)\cite{s1,s2}. For simplicity, this mechanism is demonstrated in this work by simply adding capacitors with varying capacitance values. 

The tag is designed so that both the encoded ID and sensing information are carried within the ultra-wideband (UWB) frequency of 3.1 to 10.6 GHz. The data encoding of the tag ID (ranging from '000' (ID 1) to '111' (ID 7)) is encoded within the frequency range of 3.1 to 7 GHz, while the remaining frequencies of 7 to 10.6 GHz are used to encode the sensing information. For each tag ID, three capacitance values of 0.1, 0.3, and 0.8 pF are used, demonstrating three sensing states. An example of the tag's RCS response is illustrated in Fig. \ref{tag}(b), showing the case of tag '111' (ID 7) with three capacitance values. The example illustrates that, in this work, the  variation in amplitude is utilized for detecting different IDs, while the frequency shift indicates the sensing information. 
\vspace{-1mm}
\subsection{Robot-Based Automated Data Collection}
\begin{figure*}[t]
  \begin{center}
  \includegraphics[width= 6.7in]{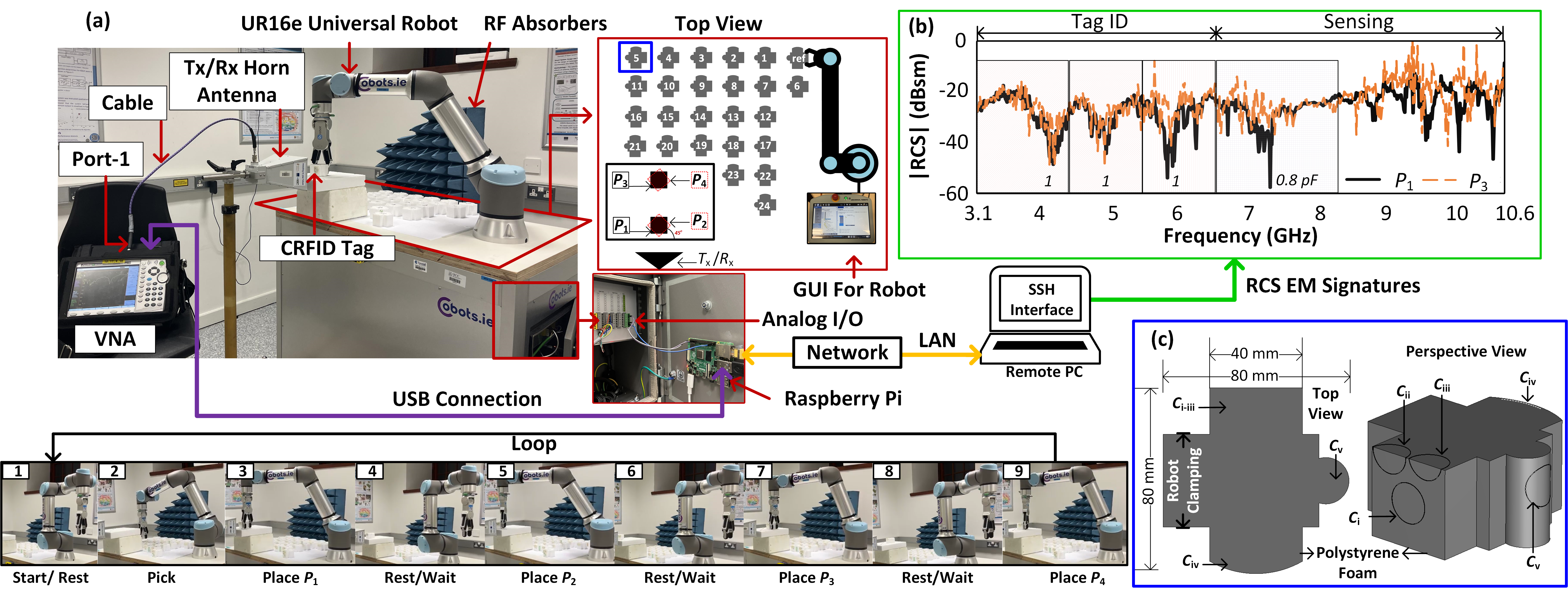}\\
 \vspace*{-2mm}
 \caption{(a) Automated RCS data acquisition setup. (b) Measured RCS of tag ID 7 with 0.8 pF sensing value at $P_{1}$ and $P_{3}$. (c) Polystyrene-made platform for mounting the tag under investigation.}
  \label{Robot}
  \end{center}
  \vspace*{-6mm}
  \end{figure*}

\vspace{-1mm}
For the purpose of collecting data for this investigation, the system depicted in Fig. \ref{Robot} was developed. All twenty-four possible tag combinations were evaluated. For each combination, data was collected for tags measured from four different positions ($\textit{P}_1$-$\textit{P}_4$) and five distinct mounting platforms (i.e., cases $\textit{C}_i$-$\textit{C}_v$).  $\textit{P}_1$ and $\textit{P}_3$ correspond to the two read ranges (200 and 300 mm), while $\textit{P}_2$ and $\textit{P}_4$ correspond, respectively, to the 45-degree tilt at these two read ranges. The variation of the mounting platform reflects five variations in the tag's physical deformation. This was facilitated by a customized polystyrene foam structure (see Fig. \ref{Robot}(c)). The flat side of the structure was utilized for the case of the tag mounted on a flat surface ($\textit{C}_i$). The edges of the same side were utilized for the cases of the tag corner bent with a length ratio of 50:50 ($\textit{C}_{ii}$) and 25:75 ($\textit{C}_{iii}$). Additionally, the structure includes two round surfaces with radii of 40 mm and 10 mm, respectively. These surfaces were used for the cases of the tag cylindrically bent ($\textit{C}_{iv}$ and $\textit{C}_{v}$). Further, to obtain the RCS of the tags, the experimental setup shown in Fig. \ref{Robot}(a) was utilized, consisting of an Anritsu MS2038C Vector Network Analyzer (VNA) interfaced with a Raspberry Pi (RPi) module using a Secure Shell (SSH) interface and the programming language Python. Libraries such as PyVISA, NumPy, and Pandas were utilized for communication and computations. The port 1 of the VNA was connected to a Schwarzbeck BBHA 9120 D - Double-Ridged Broadband Horn Antenna for transmitting and receiving EM waves \cite{19}. The VNA was used to generate a frequency sweep from 3.1 to 10.6 GHz with 0 dBm RF output power and to record the input reflection coefficient ($S_{11}$) at Port 1. The calibration method derived from \cite{20} was used to obtain the tag's RCS response from the measured $S_{11}$. 

The S-parameter (.S2P) files were acquired from the VNA with a sampling size of 700 data points once secure communication was established. The monostatic RCS was calculated using the open-source Python package Scikit-RF, following the formula in Eq.(\ref{RCSe}).
\vspace{-2mm}

\vspace{-1mm}
\begin{equation}
\label{RCSe}
\text{RCS}(\sigma) = \left| \frac{{(S_{11_{\text{Tag}}} - S_{11_{\text{Iso}}})}}{{(S_{11_{\text{Ref}}} - S_{11_{\text{Iso}}})}} \right|^2 \times \sigma_{\text{Ref}} \, \text{m}^2
\end{equation}
where $S_{11_{\text{Tag}}}$, $S_{11_{\text{Iso}}}$, and $S_{11_{\text{Ref}}}$ represent the scattering parameters of the tag, isolation, and reference, respectively.
The known RCS of the reference flat metal plate for calibration ($\sigma_{\text{Ref}}$) was calculated using Eq. (2) below,
\begin{equation}
\label{eq:RCSref}
\sigma_{\text{Ref}} = \frac{{4\pi A^2}}{{\lambda^2}} \, \text{m}^2
\end{equation}
where A is the area of the plate, and {$\lambda$} is the wavelength.

\begin{figure}[t]
  \begin{center}
  \includegraphics[width=2.9in]{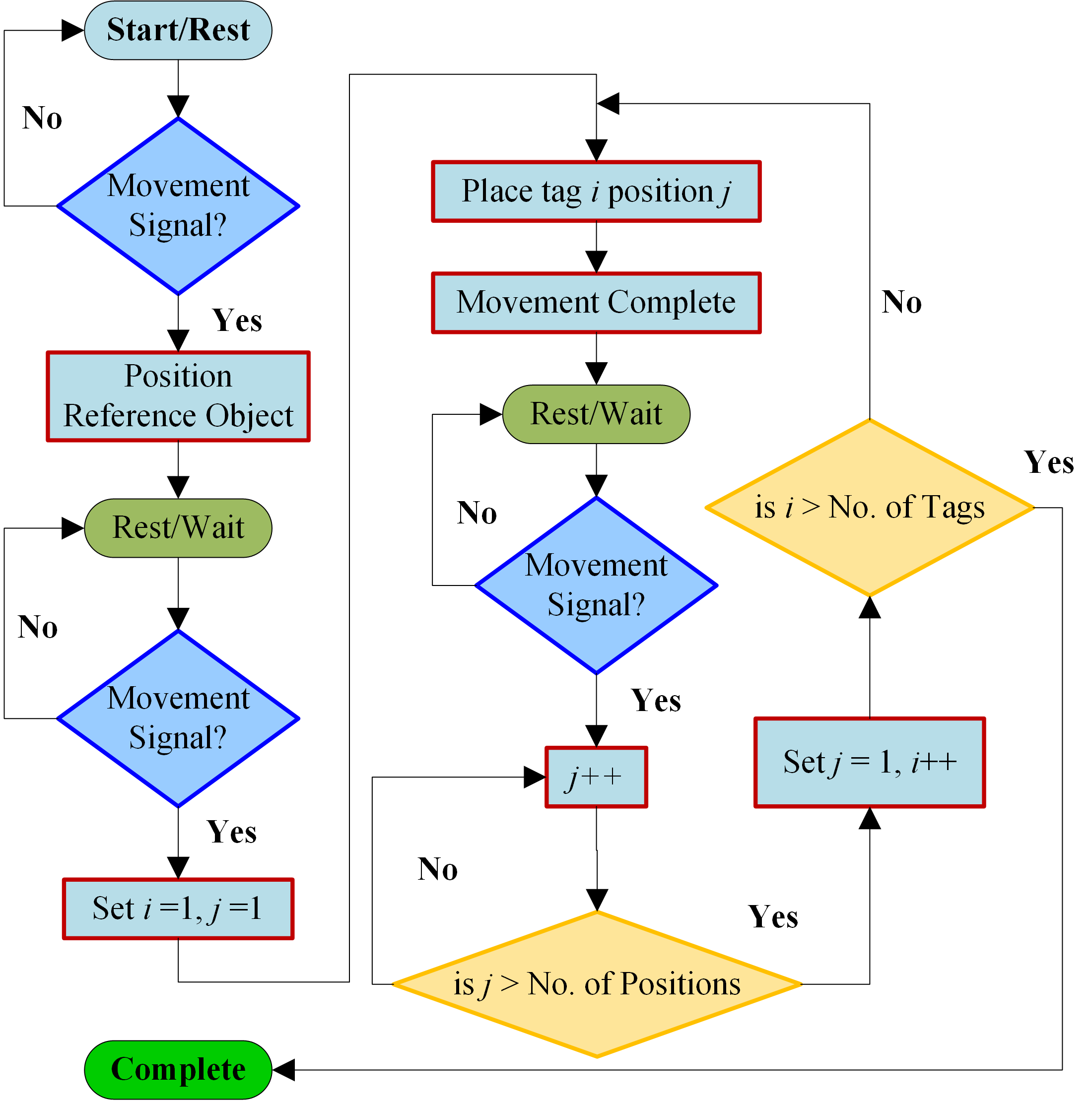}\\
 \vspace*{-1mm}
 \caption{The robot's operation flow during the pick-and-place of platform-mounted tags.}
  \label{flow1}
  \end{center}
  \vspace{-8mm}
\end{figure}

The flow of the data acquisition is explained as follows. First, the twenty-four tags were each manually mounted on the same side of the twenty-four foams (e.g., $\textit{C}_{i}$). The pick-and-place process of the foams (with mounted tags) was automated using the Universal Robot (UR16e), which has six rotating joints, a maximum move speed of 120-180 degrees per second, and a repeatability tolerance of ± 0.05 mm. The UR16e robot was programmed using Universal Robot Script (URScript), a language specific to Universal Robots, following the flowchart shown in Fig. \ref{flow1}. Initially, the robot is at the rest/wait position, and a measurement with all the structures in place with the exception of the tag was undertaken to record $S_{11_{\text{Iso}}}$. Further, the robot receives a movement signal to pick and place the foam structure with the reference calibration plate (attached to its flat side). A square flat copper plate (25 mm $\times$ 25 mm) was utilized for calibration purposes. The robot arm then returns to the rest/wait position, and measurement $S_{11_{\text{Ref}}}$ is taken. The robot is then instructed with a second movement signal to return the foam with the reference plate to its original position, followed by the pick, place ($\textit{P}_{1}$-$\textit{P}_{4}$), record $S_{11_{\text{Tag}}}$, and return of the tags \#1 to \#24. The first two calibration steps ($S_{11_{\text{Iso}}}$ and $S_{11_{\text{Ref}}}$) were performed once, followed by the RCS calculations of all twenty-four tags. Further details on the process are available in \cite{rather}. Upon the completion of $\textit{C}_{i}$, the tags were manually mounted to different sides of the foam, and the same process was repeated. \\
For each measurement scenario, twenty readings were recorded to account for varying noise levels and interferences. This yields a total dataset of 9,600 EM signatures (i.e., 24 tags x 4 positions x 5 shapes x 20 readings). In Fig. \ref{Robot}(b),  a representative example of the collected measured signatures is shown for tag ID 7, with a capacitance value of 0.8 pF measured at positions $\textit{P}_{1}$ and $\textit{P}_{3}$, displaying different levels of noise. 


\vspace{-2mm}
\section{Machine Learning and Deep Learning Modelling}
\vspace{-2mm}
Supervised ML/DL involves training models using labelled datasets to make predictions or classifications based on the provided labels. For our dataset, which requires the prediction of numerical values from given inputs instead of classes, we employ regression models. We assess the performance and suitability of four popular ML regression models: Support Vector Regression (SVR), Decision Trees (DT), Gradient Boosted Trees (GBT), and Random Forest (RF). In addition, we also explore the application of DL architectures, focusing in particular on the use of 1D CNNs in a regression-based approach.\par 
1D CNNs offer a potent approach for extracting intricate patterns and relationships from complex time-series datasets. By incorporating a 1D CNN into our analysis, we aim to take advantage of its ability to capture relevant features and enhance the predictive capabilities of our models. 
\vspace{-2mm}
\subsection{Machine Learning Based Models}
\begin{figure}[t]
  \begin{center}
  \includegraphics[width=2.7in]{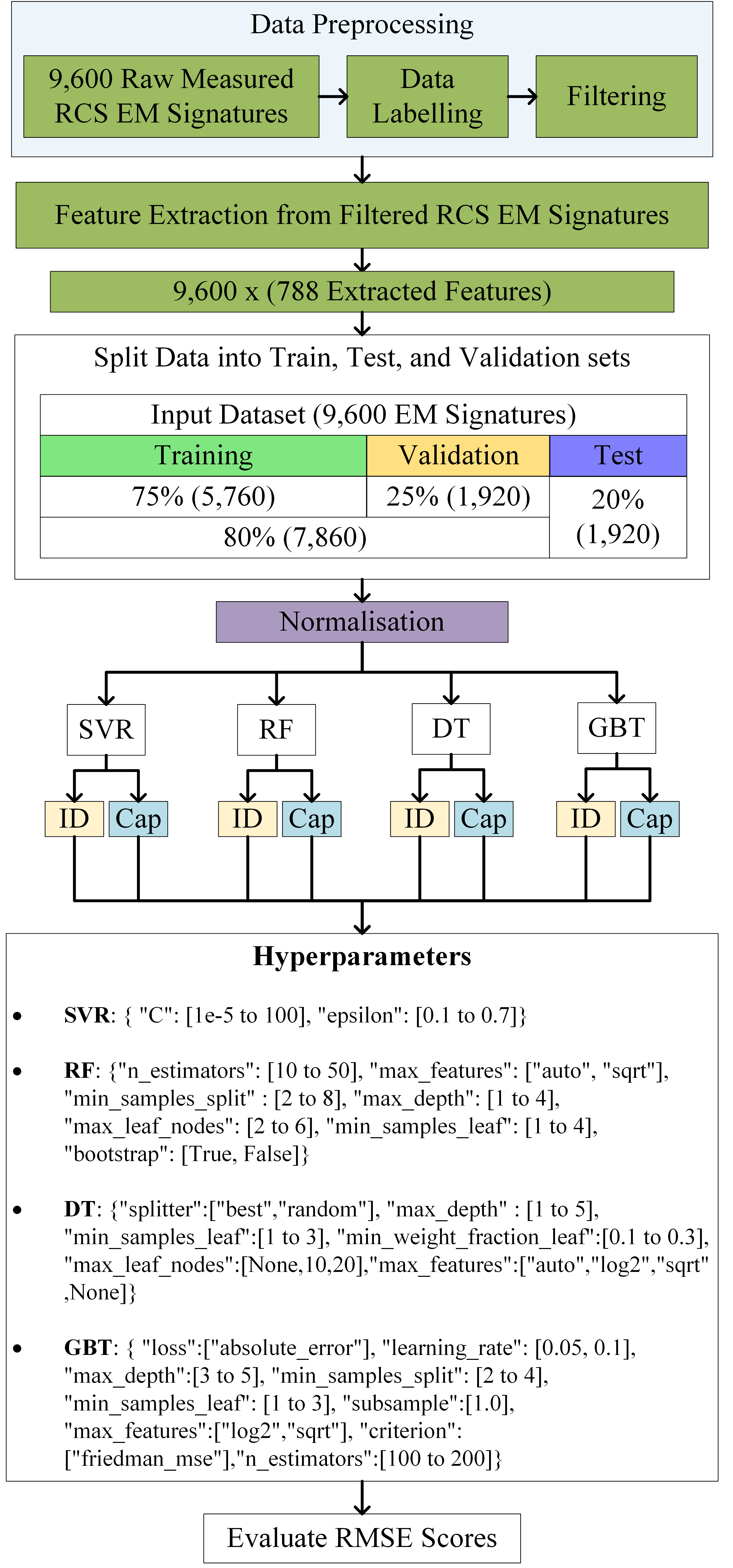}\\
 \vspace*{-1mm}
 \caption{ML implementation process flow with hyperparameters.}
  \label{flowm}
  \end{center}
  \vspace*{-8mm}
\end{figure}

\vspace{-1mm}
Once the dataset was acquired, a series of pre-processing steps were applied to the raw dataset in order to implement the ML models. The flow of the procedure used to implement ML models is shown in Fig. \ref{flowm}. Initially, the raw RCS signatures were low-pass filtered using a fourth-order Butterworth filter with a cut-off frequency of 0.1 Hz, effectively reducing noise while retaining RCS EM signatures used for encoding. This was performed to smoothen the EM signatures and reduce unwanted noise in the signal. Additionally, the appropriate output labels were added to the dataset. In regression-based ML, extracting relevant features from the raw dataset is a critical step for attaining accurate predictions. \par
To streamline this process, a TSFEL (Time Series Feature Extraction Library) was employed to extract a wide range of informative features from the given EM signatures \cite{21}. By utilizing TSFEL, the feature extraction task becomes more efficient and comprehensive. It offers a collection of built-in algorithms for capturing the essential characteristics of one-dimensional data, including statistical measures, signal processing techniques, and information theory-based features. A total of 780 features were automatically extracted using this method. These features were acquired from the full frequency band ranging from 3.1 to 10.6 GHz, which covered a total of 700 data samples of the EM signatures.
Furthermore, we included manually extracted features from specific frequency bands where relevant minima were to be found, along with their respective RCS magnitude values. We divided the data into four windows: the first window spanned from 3.1 to 4.2 GHz, the second window spanned from 4.2 to 5.2 GHz, and the third window spanned from 5.2 to 6.3 GHz. These windows were utilized to capture crucial information related to the detection of ID minima. The fourth window covered the frequency range from 6.3 GHz to 10.6 GHz and was focused for extracting features associated with sensing information. Utilizing these methods, a total of 788 features were extracted. \par
After filtering and feature extraction, the dataset was split into training, validation, and test sets. The split was stratified to ensure that the distribution of labels across the different cases remained balanced within all subsets. The data was initially split into a test set, which comprised 20\% of the original data, with the remaining 80\% of the data further divided into training and validation sets with a split ratio of 75:25. A normalization method, specifically the ’StandardScaler’ from the scikit-learn library, was then applied to the training dataset. The mean and standard deviation obtained from the training set was then used to transform the validation and test sets. For the ML model implementation, we employed a comprehensive pipeline approach to train and compare the aforementioned regression models. The pipeline encompassed several key steps. 
Firstly, we applied feature selection using Recursive Feature Elimination with Cross-Validation (RFE-CV) for each regression model. RFE-CV performed iterative feature elimination based on their impact on the model's performance, while cross-validation was utilised to ensure robustness of the feature selection process. Following feature selection, we conducted hyper-parameter optimization using GridSearchCV. The list of hyper-parameters examined for optimization is available in Fig. \ref{flowm}.\par 
The models were fine-tuned for best performance and generalization using hyper-parameter optimisation. Following our investigations in \cite{22}, each ML model was implemented for ID and sensing (capacitance values in pF) predictions separately. Each model was first trained using the training set (X\_train, y\_train), and subsequently, predictions were made on the validation set ({X\_val}, y\_val) to ensure no overfitting. If no overfitting was detected, the defined model was set and trained on the combined (X\_dev, y\_dev) training and validation sets (80\%), then evaluated on the 20\% test data (X\_test, y\_test). The root mean squared error (RMSE) was utilized as an evaluation metric to quantify the prediction performance of each regression model. By employing this detailed pipeline approach, we aimed to compare the performance of the four regression models in terms of their ability to accurately predict the ID and sensed capacitance value of the target tag. The goal was to identify the best-performing model and capitalize on its strengths to increase prediction accuracy and generalization capability in the given task. 

\vspace{-2mm}
\subsection{Deep Learning Architectures}
\vspace{-2mm}
To explore the potential of DL models, we implemented four distinct 1D CNN architectures, each tailored for ID and capacitance value prediction. These architectures use 1D CNNs to extract meaningful features from the input data and make accurate predictions. These architectures were implemented using the Keras library with TensorFlow as the backend. The model architectures consisted of multiple layers, including convolutional layers (Conv1D), pooling layers, dropout layers, batch normalization layers, and dense layers. DL models are built to automatically learn and extract hierarchical representations of the input data during the training process. \par
As a result, the model can learn and extract features directly from the raw data.  The filtered signals were thus used as the input to these models. The split and normalization procedure was kept similar to the ML models (Fig. \ref{flowm}).

\begin{figure*}[t]
  \begin{center}
  \includegraphics[width= 6in]{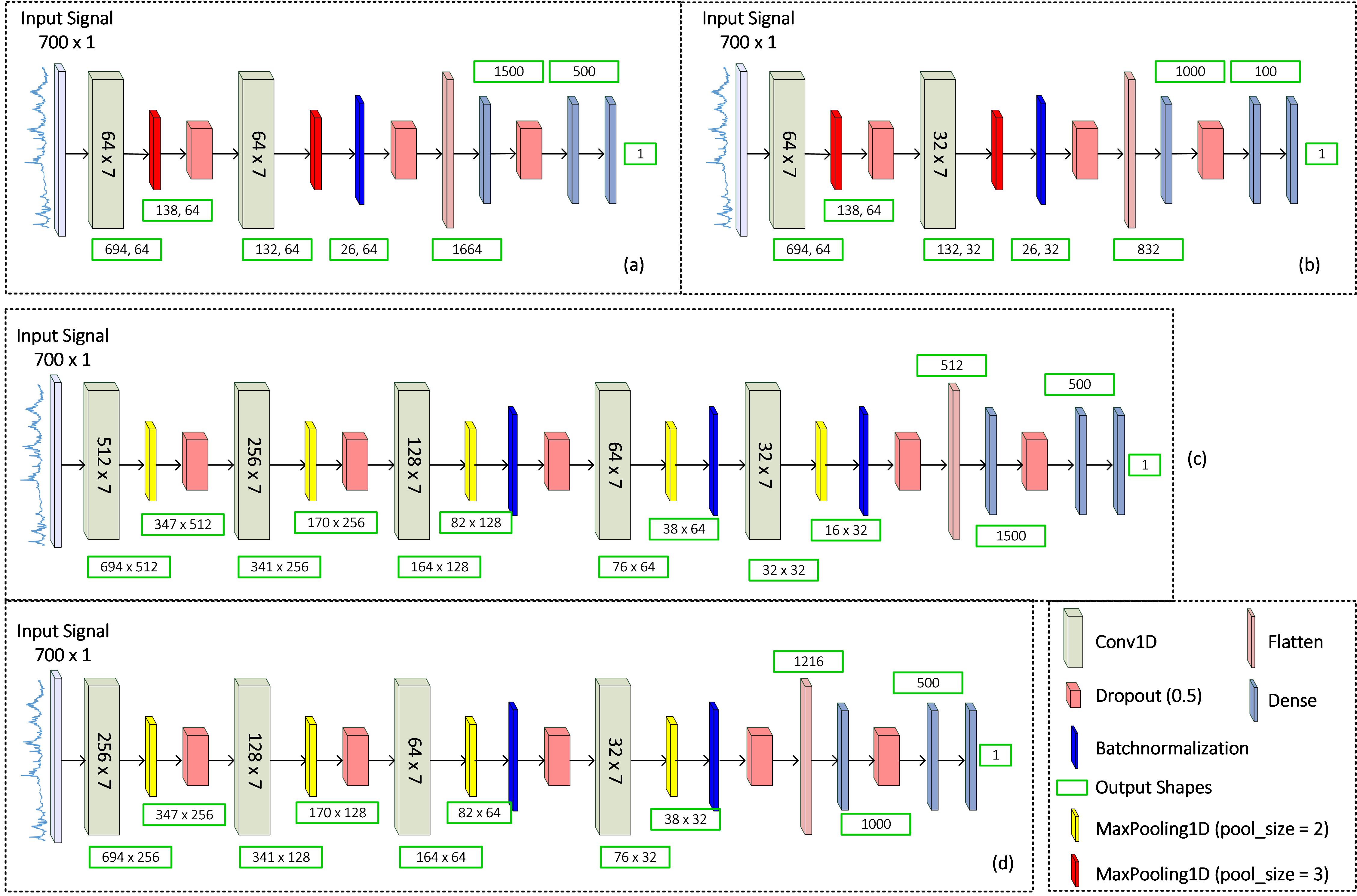}\\
 \vspace*{-2mm}
 \caption{1D CNN architectures (a) Model 1 for Tag ID, (b) Model 2 for sensing, (c) Model 3 for Tag ID, and (d) Model 4 for sensing.}
  \label{Arch}
  \end{center}
  \vspace{-6mm}
  \end{figure*} 
\vspace{-2mm}
\subsubsection{Model 1 - Tag ID}
\vspace{-2mm}
The first model (Fig. \ref{Arch}a), is intended for tag ID prediction. It consists of several layers, starting with the EM signature as the input layer. The input layer considers the data's feature dimensions, which are (700, 1), signifying 700 data points with a single feature. This is followed by a Conv1D layer with 64 filters and a kernel size of 7, which helps capture local patterns in the input data. A MaxPooling1D layer is then applied to reduce the spatial dimensions of the data. To prevent overfitting, a Dropout layer with a rate of 0.5 is included. Another Conv1D layer with 64 filters and a kernel size of 7 follows, further extracting relevant features. This is again followed by a MaxPooling1D layer and a BatchNormalization layer for normalization. Dropout is applied once more before flattening the output. The flattened data is then passed through fully-connected layers with Rectified Linear Unit (ReLU) activation functions, consisting of a Dense layer with 1,500 units, a Dropout layer with a rate of 0.5, and a Dense layer with 500 units. Finally, a Dense layer with a linear activation function is used in the output layer for regression-based tag ID prediction. 
\vspace{-2mm}
\subsubsection{Model 2 - Tag Sensing}
\vspace{-2mm}
The second model focuses on the tag capacitance value prediction (Fig. \ref{Arch}b). Similar to the first model, it starts with an input layer and then proceeds with a Conv1D layer with 64 filters and a kernel size of 7. This is followed by a MaxPooling1D layer for down-sampling and a Dropout layer for regularization. Another Conv1D layer with 32 filters and a kernel size of 7 is applied, followed by another MaxPooling1D layer and a BatchNormalization layer. Dropout is included to further prevent overfitting. The output is then flattened and passed through fully connected layers, consisting of a Dense layer with 1,000 units, a Dropout layer with a rate of 0.5, and a Dense layer with 100 units. The output layer uses a linear activation function for regression-based capacitance values prediction.
\vspace{-2mm}
\subsubsection{Model 3 - Tag ID (Extended)}
\vspace{-2mm}
The third model is an extended version of the tag ID prediction Model 1. It includes additional Conv1D layers to capture more complex patterns (Fig. \ref{Arch}c). The architecture starts with an input layer and proceeds with a Conv1D layer with 512 filters and a kernel size of 7. A MaxPooling1D layer is then applied for down-sampling, followed by Dropout for regularization. The subsequent layers consist of Conv1D layers with 256, 128, 64, and 32 filters, accompanied by a MaxPooling1D layer, BatchNormalization layer, and Dropout layer. The output is flattened and passed through fully-connected layers, including a Dense layer with 1,500 units, a Dropout layer, a Dense layer with 500 units, and a final Dense layer with a linear activation function for tag ID prediction.
\vspace{-2mm}
\subsubsection{Model 4 - Tag Sensing (Extended)}
\vspace{-2mm}
The fourth model is an extended version of the tag capacitance value prediction Model 2. It incorporates additional Conv1D layers to capture more intricate patterns in the data (Fig. \ref{Arch}d). The architecture starts with an input layer, followed by Conv1D layers with 256, 128, 64, and 32 filters, each accompanied by a MaxPooling1D layer, BatchNormalization layer, Dropout layer, and MaxPooling1D layer. The flattened output is then passed through fully-connected layers, consisting of a Dense layer with 1,000 units, a Dropout layer, a Dense layer with 500 units, and a final Dense layer with a linear activation function for capacitance prediction. No padding was used in any layer of the architecture. \par
All the discussed architectures were trained using the Adam optimizer with the mean squared error (MSE) loss function. The training process involved a batch size of 32 and was conducted over 300 epochs. Early stopping was implemented to monitor the training process and halt if the loss did not improve for a certain number of epochs. Model checkpoints were saved to retain the best-performing model. The  training loss was tracked to assess the convergence and overall performance of the models. The evaluation of the trained models was performed using the RMSE as the evaluation metric, measuring the average deviation between the predicted and actual values. Similar to the ML model implementation, the models were trained first on the training set, followed by a validation set to make sure there was no overfitting. Finally, if no overfitting was detected, the architectures were trained on the development set and tested on the unseen test dataset.   
\vspace{-2mm}
\section{Results}
\vspace{-2mm}
A summary of RMSE results from the ML and DL models is shown in Table \ref{tab}.  For all investigated models, the evaluation results on the training, validation, and test sets are given. It is seen that both ML models and 1D CNN architectures were able to generalise well with the data to make predictions on ID and capacitance values. 
As shown, the GBT model achieved the best performance with an RMSE of 0.3 (normalized RMSE: 4.2\%) on the test set for tag ID prediction, while its sensing prediction RMSE was relatively high with 0.127 (18.4\%). SVR performed slightly better as compared to the DT and RF in detecting both ID and capacitance values. However, all these models had a significantly high RMSE for the capacitance value prediction.  Moving on to the DL architectures, for tag ID prediction, 1D CNN Model 1 achieved a low RMSE of 0.098 (1.2\%) on the test set. In terms of sensing prediction, the RMSE was 0.0304 (4.3\%). These are significantly better than the best-performing GBT model, with a 67.33\% and 73.23\% decrease in RMSE error for tag ID and sensing values, respectively. \par

\begin{table}[!htb]
    \centering
    \renewcommand{\arraystretch}{1.2}
    \caption{Summary of RMSE Results}
    \label{tab}
    \begin{tabular}{c|c|c|c|c|c|c}
        \hline
        \textbf{Model} & \multicolumn{3}{c|}{\textbf{Tag ID 0-7}} & \multicolumn{3}{c}{\textbf{Sensing (0.1 pF - 0.8 pF)}}\\
        \hline
        \hline
         & Train & Val* & Test & Train & Val* & Test \\
        \hline
        SVR & 0.612 & 0.616 & 0.600 & 0.226 & 0.233 & 0.227 \\
        \hline
        GBT & 0.249 & 0.315 & 0.300 & 0.114 & 0.137 & 0.127 \\
        \hline
        DT & 1.122 & 1.114 & 1.076 & 0.271 & 0.273 & 0.269 \\
        \hline
        RF & 0.906 & 0.889 & 0.916 & 0.260 & 0.261 & 0.259 \\
        \hline
        1D CNN-1,2 & 0.108 & 0.086 & 0.098 & 0.026 & 0.031 & 0.0304 \\
        \hline
        \textbf{1D CNN-3,4} & \textbf{0.073} & \textbf{0.084} & \textbf{0.061} & \textbf{0.019} & \textbf{0.030} & \textbf{0.0241} \\
        \hline
    \end{tabular}\\
    \footnotesize\textit{Best performing model is highlighted in bold (Val* = Validation)}

\end{table}

\begin{figure}[t]
  \begin{center}
  \includegraphics[width=2.8in]{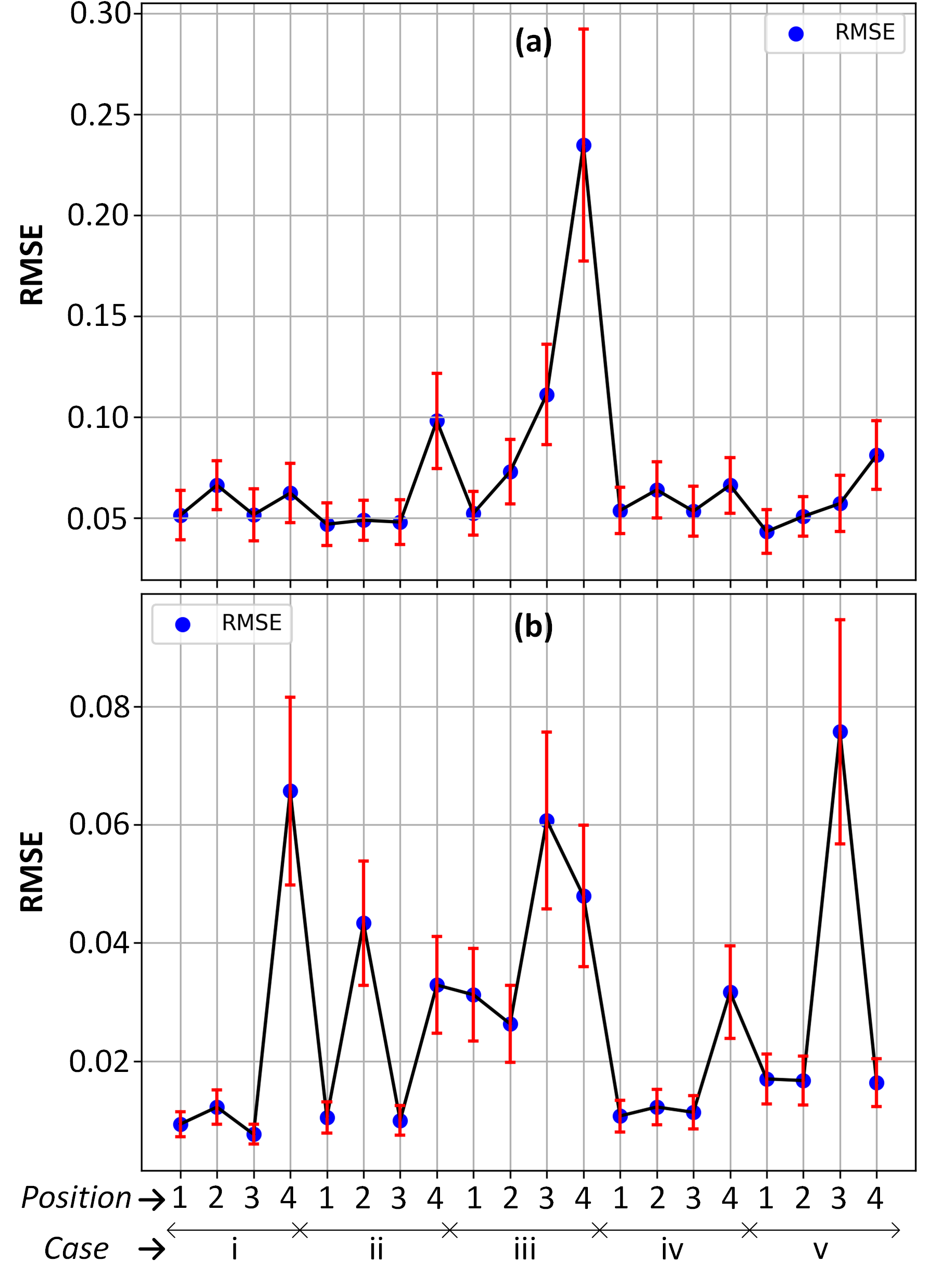}\\
 \vspace*{-2mm}
 \caption{RMSE and standard deviation results from different tag positions (i.e., $\textit{P}_1$-$\textit{P}_4$) and tag deformations (i.e., $\textit{C}_i$-$\textit{C}_v$) for: (a) Tag ID using Model 1 and (b) Sensing using Model 2.}
  \label{r1}
  \end{center}
  \vspace*{-8mm}
\end{figure}

\begin{figure}[t]
  \begin{center}
  \includegraphics[width=2.8in]{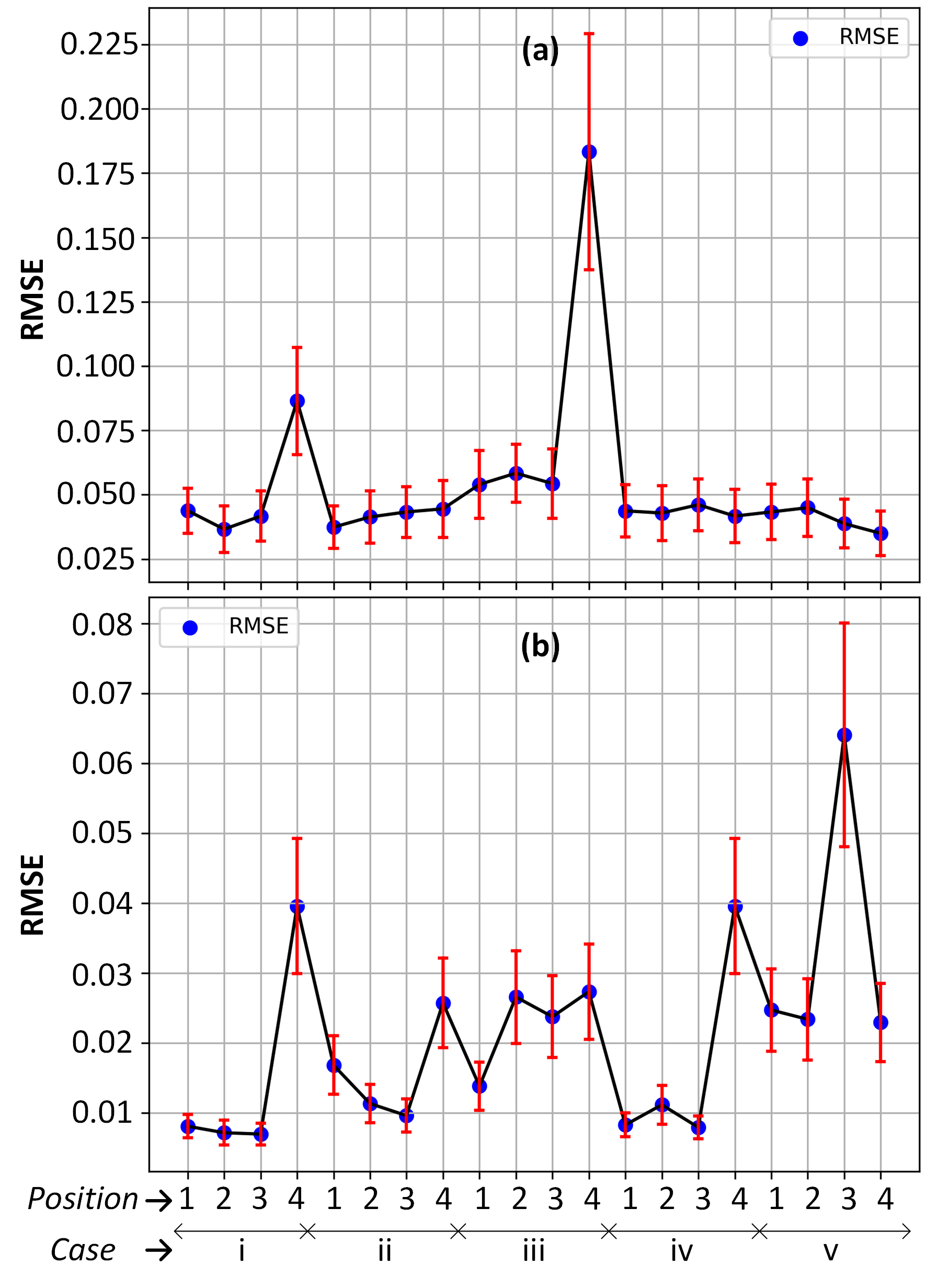}\\
 \vspace*{-1mm}
 \caption{RMSE and standard deviation results from different tag positions (i.e., $\textit{P}_1$-$\textit{P}_4$) and tag deformations (i.e., $\textit{C}_i$-$\textit{C}_v$) for: (a) Tag ID using Model 3 and (b) Sensing using Model 4.}
  \label{r2}
  \end{center}
  \vspace*{-5mm}
\end{figure}

The deeper 1D CNN architectures (Models 3 and 4) achieved even better results with an RMSE of 0.061 (0.87\%) and 0.0241 (3.44\%) for ID and sensing, respectively. Compared to 1D CNN Models 1 and 2, the RMSE was reduced by 37.76\% and 20.78\%, respectively.
In summary, among the ML models, GBT exhibited the lowest RMSE for both tag ID and capacitance value prediction tasks. On the other hand, the 1D CNN architectures remarkably outperformed the ML models. These findings suggest that DL approaches using 1D CNN have the potential to provide valuable insights and accurate predictions in this domain compared to classical ML models based on feature extraction. For a more detailed analysis of the achieved results, the four 1D CNN models are further evaluated for all 20 cases (i.e., 4 positions $\times$ 5 shapes). The RMSE for tag ID and capacitance value prediction at different cases and positions as obtained from Models 1 and 2, respectively, is depicted in Fig. \ref{r1}. 
It is seen that RMSE of the Model 1 is consistent for all cases except for $\textit{C}_{iii}$, at the longer read range with 45 degrees tilt (${P}_4$) scenario. Similarly, for capacitance sensing value, the prediction shows that the RMSE is slightly higher at the ${P}_4$ positions. Furthermore, in Models 3 and 4, the RMSE shows similar trends (as shown in Fig. \ref{r2}) with slightly more consistent and lower values of RMSE. The capacitance-sensing Model 4 also shows a reduction in RMSE for all cases. 
\par
Models 3 and 4 were the best performing models that were saved to then make predictions in a real-world scenario. To validate the effectiveness of these models, we conducted experiments with an Anritsu MS2038C VNA connected to a horn antenna \cite{19}, and interfaced with a RPi. The raw data obtained from the VNA was processed using a custom-written program that employed the RCS computation and the trained Models 3 and 4 for tag ID and capacitance value predictions, respectively. Ten (2\%) of the total 480 possible scenarios (24 Tags $\times$ 4 Positions $\times$ 5 Cases ) were selected at random and utilized for the validation. As shown in Table \ref{validation}, it is observed predictions made for ID and capacitance sensing values were generally accurate for different testing scenarios. In practical scenarios, the predicted ID from the regression model would be rounded to the nearest integer. For instance, an ID predicted as 2.7 would be regarded as 3. By employing this rounding method, Table II demonstrates that the ID was predicted correctly in every instance. Similarly, rounding the capacitance values to the nearest value (0.1, 0.3, or 0.8 pF) resulted in accurate capacitance sensing value predictions in eight out of ten instances. The remaining two errors were due to the proximity between the rounded values of 0.1 and 0.3. The variability in errors can also be attributed to the variation in the model's RMSE for different shapes and positions. The measurement process and an example of the tag ID and capacitance sensing value detection using the trained models are available in a demonstration video, that can be viewed in \cite{demo}.

\begin{table}[!htb]
    \centering
    \renewcommand{\arraystretch}{1.2}
    \caption{Actual vs Predicted Values with Errors}
    \label{validation}
    \begin{tabular}{cccccc}
    \hline
    \textbf{$\textit{C}_{i-v}$ ($\textit{P}_{1-4}$)} & \textbf{$\textit{ID}_{act.}$} & \textbf{$\textit{S}_{act.}$} & \textbf{$\textit{ID}_{pred.}$ (Error)} & \textbf{$\textit{S}_{pred.}$ (Error)} \\
    \hline
    \hline
    ii (${P}_2$) & 3 & 0.8 & 2.8981 (0.1019) & 0.7360 (0.0640) \\
    v (${P}_1$) & 7 & 0.3 & 6.9721 (0.0279) & 0.2967 (0.0033) \\
    i (${P}_4$) & 5 & 0.3 & 4.9641 (0.0359) & 0.2640 (0.0360) \\
    iii (${P}_3$) & 1 & 0.1 & 1.0199 (0.0199) & 0.1647 (0.0647) \\
    iv (${P}_1$) & 2 & 0.8 & 2.0138 (0.0138) & 0.7599 (0.0401) \\
    i (${P}_2$) & 5 & 0.3 & 4.8707 (0.2293) & 0.3138 (0.0138) \\
    v (${P}_3$) & 6 & 0.3 & 6.1629 (0.1629) & 0.2733 (0.0267) \\
    iii (${P}_4$) & 3 & 0.1 & 2.8091 (0.1909) & 0.1567 (0.0567) \\
    i (${P}_2$) & 2 & 0.3 & 2.0148 (0.0148) & 0.2959 (0.0041) \\
    iv (${P}_1$) & 7 & 0.3 & 6.6830 (0.3170) & 0.2961 (0.0039) \\
    \hline
  \end{tabular}\\
  \footnotesize\textit{$act.$ = Actual, $pred.$ = Predicted, $S$ = Capacitance Value (pF)}
  \vspace{-5mm}

\end{table}
\vspace{-3mm}
\section{Discussion} 
\vspace{-3mm}
The presented study provides a comprehensive analysis of ML and DL models for tag ID and sensing prediction in CRFID systems. The results demonstrate the effectiveness of both ML and DL approaches, with DL models exhibiting superior performance in terms of accurately predicting tag ID and sensed capacitance values. The findings highlight the potential of DL models, particularly 1D CNN architectures, in providing valuable insights and accurate predictions in the CRFID domain compared to classical ML models based on feature extraction.
The utilization of a large dataset acquired through an automated data acquisition system sets this study apart from previous research works, which are limited in their dataset size due to the constraints of data collection procedures \cite{14,15,16}. The comprehensive dataset enables the development of robust ML/DL models that can learn and adapt to varying EM signatures which encode data. The automation helps to incorporate varying scenarios for the CRFID tags, improving the robustness and reliability of the proposed models and enhancing their practical applicability. Investigating the impacts of varying tag surface shapes, orientations, and read ranges is paramount for real-world implementation. CRFID tags may encounter different surface shapes and orientations in practical scenarios, and the read ranges may vary. The developed detection algorithms can be optimized by studying these factors to handle diverse environmental conditions and ensure reliable tag detection.\par
Furthermore, the feasibility of extracting both ID and sensing information from the EM signatures is another significant contribution of this study. Traditionally, CRFID systems have focused primarily on tag identification \cite{14,15}. However, by exploring the extraction of sensing information from the EM signatures, this research paves the way for leveraging CRFID technology in applications that require continuous monitoring and sensing capabilities (For example, for food quality monitoring, health monitoring,   etc. \cite{karm}), expanding the potential applications of CRFID systems beyond simple identification tasks. The study also investigates regression-based approaches for accurate and continuous prediction of tag IDs and sensed capacitance values. This regression-based approach enables precise and continuous prediction of tag information, which is particularly useful in applications where real-time and continuous monitoring is crucial. The results obtained from these regression models provide insights into the potential accuracy and reliability of CRFID systems in predicting tag IDs and sensing values. Furthermore, this paper aims to highlight one of the possible future impacts of integrating ML/DL with RFID and antenna technology, with other promising applications being investigated for antenna design optimisations, RF signal processing etc. in \cite{j,k,l}. While ML/DL has not yet made significant strides in this field, we anticipate that this will change soon, paving the way for advancements in this evolving field and ultimately transforming industries such as supply chain management, logistics, health monitoring, chipless and sustainable Identification, and sensing technologies. 

There are, however, certain limitations and challenges that need to be acknowledged. Firstly, the effectiveness of the proposed algorithms and models may be influenced by the specific hardware configurations and antenna designs. The customization and optimization efforts required for different hardware configurations and encoding techniques should be considered for practical implementation.
\vspace{-3mm}
\section{Conclusion}
\vspace{-3mm}
ML and DL models were implemented in this paper for a 3-bit CRFID sensor tag. For the first time, an automated data acquisition methodology was utilized to collect a large dataset to implement these models incorporating varying surface shapes, read ranges, and tilt angles for robust detection. A thorough analysis of four popular ML models, namely SVR, GBT, DT, and RF was carried out. Furthermore, 1D CNN architectures were applied for our analysis as well. It was found that the GBT performed well in the ML models; however, CNN models outperformed all ML models with impressive low RMSE of 0.061 (0.87\%) and 0.0241 (3.44\%) for ID and sensing, respectively. The 1D CNN architectures thus display an impressive outcome in generalizing well to the given EM signatures with varying parameters. The best models were used to infer predictions in real-world scenarios with successful results. In future, our goal is to further enhance those models and lower the RMSE in a wide range of cases while also developing models for higher bit capacity tags. 
\vspace{-3mm}
\section*{Acknowledgments}
\vspace{-3mm}
This paper has emanated from research funding provided by Science Foundation Ireland (SFI) as part of the SFI Centre VistaMilk (SFI 16/RC/3835). Aspects of this work have been supported under the following Grant Numbers: CONNECT Centre for Future Networks and Communications (13/RC/2077), Insight Centre for Data Analytics (SFI/12/RC/2289) and 16/RC/3918-CONFIRM, as well as the Enterprise Ireland funded Holistics DTIF (Disruptive Technologies Innovation Fund) (EI-DT20180291-A). All work is co-funded under the European Regional Development Fund.
\vspace{-3mm}

\end{document}